# Variable-Rate *M*-PSK Communications without Channel Amplitude Estimation

Athanasios S. Lioumpas, *Student Member, IEEE,* and

George K. Karagiannidis, *Senior Member, IEEE*

## Abstract

Channel estimation at the receiver side is essential to adaptive modulation schemes, prohibiting low complexity systems from using variable rate and/or variable power transmissions. Towards providing a solution to this problem, we introduce a variable-rate (*VR*) *M*-PSK modulation scheme, for communications over fading channels, in the absence of channel gain estimation at the receiver. The choice of the constellation size is based on the signal-plus-noise (*S+N*) sampling value rather than on the signal-to-noise ratio (*S/N*). It is analytically shown that *S+N* can serve as an excellent simpler criterion, alternative to *S/N*, for determining the modulation order in *VR* systems. In this way, low complexity transceivers can use *VR* transmissions in order to increase their spectral efficiency under an error performance constraint. As an application, we utilize the proposed *VR* modulation scheme in equal gain combining (EGC) diversity receivers.

The authors are with the Division of Telecommunications, Electrical and Computer Engineering Department, Aristotle University of Thessaloniki, 54124 Thessaloniki, Greece (e-mail: {alioumpa;geokarag}@auth.gr).

# I. INTRODUCTION

A common technique for coping with fading in wireless communications, is transmission or reception diversity, which provides performance improvement at the cost of extra hardware or increased bandwidth usage. Alternatively, if a feedback link is available, the fading can be mitigated by allowing the receiver to monitor the channel conditions and request compensatory changes in certain parameters of the transmitted signal. This technique is called adaptive transmission and its basic concept is the real-time balancing of the link budget through adaptive variation of the transmitted power level, symbol transmission rate, constellation size, coding rate/scheme, or any combination of these parameters.

Adapting to the signal fading allows the channel to be used more efficiently since power and rate can be allocated to take advantage of favorable channel conditions. Adaptive transmission was first proposed by Hayes [1], who considered a Rayleigh fading channel where the amplitude of the transmitted signal is under control of the receiver through a feedback channel.

## A. Motivation

Several works have dealt with adaptive transmission systems, in which the modulation/coding parameters, such as constellation size and coding rate are adaptively varied with the prevailing fading channel condition while maintaining a certain error rate performance [2]-[9]. These techniques have been also considered in several modern terrestrial and satellite communication systems, also incorporated in standards (e.g. High-Speed Downlink Packet Access - HSDPA [12] and Digital Video Broadcasting - Return Channel via Satellite - DVB-RCS [13]).

In many practical applications, the adaptive transmission is limited to variable-rate (*VR*) schemes, since the difference from variable-power (*VP*) *VR*, is a small fraction of decibel for most types of fading [4]. Most of the *VR* communication systems proposed in the literature, adapt the transmission parameters according to the instantaneous channel condition, limiting their applicability to systems that estimate the instantaneous channel gain, either perfectly or with estimation errors.

## B. Previous Works and Contribution

To the best of the authors' knowledge, no *VR* modulation scheme has been proposed in the literature that does not require channel gain estimation, neither at the receiver, nor at the transmitter. In [14], the performance of a blind adaptive modulation scheme is investigated, which necessitates channel knowledge only at the receiver. In this scheme, the order of an $M$-ary quadrature amplitude modulation ($M$-QAM) sequentially decreases or increases according to positive/negative acknowledgments (ACK/NACK) sent to the transmitter.

In this paper, we propose a rate adaptation technique, which is based on the signal-plus-noise ($S+N$) samples, instead of the instantaneous $S/N$. Owing to this particularity, no channel gain estimation is necessary for determining the transmission rate.

The $S+N$ technique has been applied to selection diversity communication systems (e.g. see [15] and the references therein), where its main advantage lies in taking into account the instantaneous noise power, in contrast to $S/N$ (e.g. the channel gain divided by the average noise power). The main contribution of this paper can be summarized as follows.

- We analytically show that $S+N$ can serve as an excellent alternative criterion to $S/N$, in $VR$ systems; the same spectral efficiency can be achieved, when either $S/N$ or $S+N$ is used for determining the constellation size.

- Based on this important outcome, we introduce an $S+N$-based $VR$ $M$-ary phase shift keying ($M$-PSK) modulation scheme that does not require any channel gain estimation. Note, that $M$-PSK modulation schemes have some important advantages over QAM schemes, e.g. improved performance when used with high power amplifiers (HPA). Another advantage of using $M$-PSK instead of $M$-QAM for Orthogonal Frequency-Division Multiplexing (OFDM) systems is the availability of techniques to reduce the signal's Peak-to-Average-power Ratio (PAR), also known as PAPR or PMEPR (Peak-to-Mean Envelope Power Ratio) [16].

*C. Outline*

The remainder of the paper is organized as follows. In Section II, the $S+N$ based adaptation scheme is presented and compared with the $S/N$ one. Two criteria are applied for evaluating the efficiency of the $S+N$ scheme. The $VR$ $M$-PSK modulation scheme is presented in Section III, along with an application to EGC receivers. Section IV discusses some numerical results. Concluding remarks are given in Section V.

## II. SYSTEM MODEL AND *VR* SCHEMES

We consider a discrete-time channel, assuming that the fading amplitude $a[i]$ follows a Rayleigh distribution with probability density function (pdf)

$$f_a(a) = \frac{2a}{\Omega} e^{-\frac{a^2}{\Omega}}, \tag{1}$$

where $\Omega = E\left\{a^2\right\}$. The channel gain is assumed to be the same between two symbol intervals. The instantaneous received signal is

$$r[i] = a[i] e^{j\vartheta[i]} s[i] + n[i], \tag{2}$$

where $s[i]$ is the complex transmitted symbol, $\vartheta[i]$ is the phase introduced by the fading channel and $n[i]$ is a zero mean circularly symmetric complex Gaussian noise with $E\{n[i]^* n[i]\} = N_0 = 2\sigma^2$.

The symbol $s[i]$ is selected from a signal constellation with $M$ symbols (e.g. *M*-QAM or *M*-PSK) each with energy $E_{S_k}$ and total average symbol energy, $E_S = 1$. Then, the received *S/N* will be

$$\gamma[i] = E_S \frac{a^2[i]}{B N_0}, \tag{3}$$

where $B$ is the bandwidth of the received signal. The average channel gain is denoted as $\overline{\gamma}$, which is assumed to be $\overline{\gamma} = 1$, without loss of the generality.

### A. *S/N-based VR Systems*

In conventional *VR* systems, the modulation order used at the time instant $i + 1$, is determined by the value of the received *S/N*, $\gamma[i]$, at the time instant $i$, and a predetermined symbol or bit error rate (SER or BER) target. The objective is to maximize the spectral efficiency by using the largest possible constellation size for transmission under the instantaneous error rate requirement. Towards this aim, a mathematical formulation that relates the SER, $P_S$, with the modulation order, $M$, and the received *S/N*, $\gamma[i]$, are necessary. This formulation must be simple enough, so that $M$ can be directly expressed as a function of $P_S$ and $\gamma[i]$, i.e.,

$$P_S = G\left(\gamma[i],\, M\right) \iff M = f\left(\gamma[i],\, P_S\right) \quad \text{or } \gamma[i] = V\left(M,\, P_S\right) \tag{4}$$

In the following, the time notation, $i$, will be omitted.

Given a direct relation between $M$ and $\gamma$, at each symbol time, the constellation order, $M$, is selected from $N$ available ones, i.e., $\{M_1, M_2, M_3, M_4, \ldots, M_N\}$, depending on which region $M = f\left(\gamma,\, P_S\right)$ lies; the modulation order will be equal to $M_j$, if $M_j \leq M < M_{j+1}$, with $j = 1, \ldots,$ $N$ and $M_{N+1} = \infty$. Because of the relation between $M$, $\gamma$ and $P_S$, the constellation size can be equivalently determined by $\gamma$, after discretizing the range of channel fade levels. Specifically, the range of $\gamma$ is divided into $N + 1$ fading regions $[\gamma_j, \gamma_{j+1}]$, and the constellation $M_j$ is associated with the $j$th region.

In summary, the modulation order is determined as

$$
M = M_j, \quad \text{if} \left\{
\begin{array}{c}
M_j \leq M < M_{j+1} \\
\text{or} \\
\gamma_j \leq \gamma < \gamma_{j+1} \Leftrightarrow \gamma_j \leq E_S \frac{a^2}{N_0} < \gamma_{j+1},
\end{array}
\right. \tag{5}
$$

where $\gamma_j$s are calculated from $M = f(\gamma, P_S)$.

For example, assuming a *VR M*-QAM scheme, a mathematical expression for associating $M$, $\gamma$ and $P_S$ may be the following bound [17, Ch. 7]

$$
P_S \leq 2\text{erfc}\left( \sqrt{\frac{3\gamma}{2(M-1)}} \right), \tag{6}
$$

with erfc$(\cdot)$ denoting the complementary error function. Eq. (6) can be solved with respect to $M$, to obtain the maximum modulation order as a function of the instantaneous *S/N*, as

$$
M = 1 + \frac{3\gamma}{\left( \sqrt{2}\text{erfc}^{-1}\left( P_s/2 \right) \right)^2}, \tag{7}
$$

where erfc$^{-1}(\cdot)$ stands for the inverse complementary error function.

Finally, from (7), the fading regions are obtained as

$$
\gamma_j = \frac{(M_j - 1)}{3} \left( \sqrt{2}\text{erfc}^{-1}\left( P_s/2 \right) \right)^2 \tag{8}
$$

### B. S+N-based VR Systems

In this paper, we propose an alternative criterion for determining the modulation order in *VR* systems, which does not require the estimation of the fading amplitude and is only based on a scaled *S+N* sample as

$$
\xi = \frac{\Re\left\{ r[i]e^{-j\vartheta[i]} \right\}^2}{N_0} = \frac{\left( a\sqrt{E_{S_k}} + n_I \right)^2}{N_0} \tag{9}
$$

where $\Re\{z\}$ denotes the real part of $z$ and $n_I$ is the in-phase noise component with variance $N_0/2$.

In the proposed $S+N$ scheme, the modulation order is chosen similarly to the conventional one, with the difference that $\gamma$ is replaced by $\xi$, i.e. the selected order, $M$, will be

$$M = M_j, \quad \text{if} \quad \begin{cases} \gamma_j \leq \xi < \gamma_{j+1} \\ \text{or} \\ \gamma_j \leq \frac{\left(a\sqrt{E_{S_k}} + n_I\right)^2}{N_0} < \gamma_{j+1}, \end{cases} \quad (10)$$

The effectiveness of the proposed scheme will be compared against the conventional one by considering the *S/N* as the reference criterion for determining the maximum modulation order. The question that has to be answered is *how would the choice of the modulation order of a VR system be affected by utilizing the S+N instead of the S/N.*

In the following, assuming constant energy transmissions, the answer will be given through the calculation of two specific probabilities-criteria, which eventually quantify the similarity between $S+N$ and *S/N.*

*Criterion 1: The probability that both S/N and S+N determine the same constellation size*, i.e.,

$$\Pi_1 = \text{Pr} \left\{ \gamma_j \leq E_S \frac{a^2}{N_0} < \gamma_{j+1} \ \cap \ \gamma_j \leq \frac{\left(a\sqrt{E_S} + n_I\right)^2}{N_0} < \gamma_{j+1} \right\}. \quad (11)$$

Next, we present an approach for deriving a closed-from solution for $\Pi_1$.

By setting $y = a\sqrt{E_S}$, (11) can be equivalently re-written as

$$\Pi_1 = \text{Pr} \left\{ \sqrt{N_0 \ \gamma_j} \leq y < \sqrt{N_0 \gamma_{j+1}} \ \cap \ \sqrt{N_0 \gamma_j} - y \leq n_I < \sqrt{N_0 \gamma_{j+1}} - y \right\}, \quad (12)$$

or

$$\Pi_1 = \int\limits_{\sqrt{N_0 \ \gamma_j}}^{\sqrt{N_0 \gamma_{j+1}}} \text{Pr} \left\{ \sqrt{N_0 \gamma_j} - \ y \leq n_I \ < \ \sqrt{N_0 \gamma_{j+1}} - \ y \right\} f_y(y) dy. \quad (13)$$

The pdf of the random variable, $y$, can be directly calculated as follows, after applying the transformation [19, (5.6)] in (1),

$$f_y(y) = \frac{1}{\sqrt{E_S}} f_a\left(\frac{y}{\sqrt{E_S}}\right) = \frac{2y}{\Omega E_s} e^{-\frac{y^2}{\Omega E_s}}. \quad (14)$$

Moreover, the probability

$$\text{Pr} \left\{ \sqrt{N_0 \gamma_j} - y \leq n_I < \sqrt{N_0 \gamma_{j+1}} - y \right\} \quad (15)$$

involved in (13) is the probability that the random Gaussian variable, $n_I$, lies in a specific interval and it is directly related to the complementary error function $\text{erfc}(\cdot)$. Using the cumulative distribution function (cdf) of the zero-mean Gaussian random variable, $x$, [18, (A.18)]

$$F_X(x) = \int_{-\infty}^{x} \frac{1}{\sqrt{2\pi}\sigma} e^{-\frac{z^2}{2\sigma^2}} dz$$

and the definition of the erfc($\cdot$), [18, (A.24)]

$$\text{erfc}(x) = \frac{2}{\sqrt{\pi}} \int_{x}^{\infty} e^{-z^2} dz,$$

the probability (15) can be calculated after some trivial manipulations as

$$\Pr\left\{\sqrt{N_0\gamma_j} - y \leq n_I < \sqrt{N_0\gamma_{j+1}} - y\right\} = \frac{1}{2}\left[\text{erfc}\left(\frac{\sqrt{N_0\gamma_j} - y}{\sqrt{N_0}}\right) - \text{erfc}\left(\frac{\sqrt{N_0\gamma_{j+1}} - y}{N_0}\right)\right]. \tag{16}$$

Therefore, the probability, $\Pi_1$, can be written as

$$\Pi_1 = \frac{1}{2} \int_{\sqrt{N_0\gamma_j}}^{\sqrt{N_0\gamma_{j+1}}} \left[\text{erfc}\left(\frac{\sqrt{N_0}\gamma_j - y}{\sqrt{N_0}}\right) - \text{erfc}\left(\frac{\sqrt{N_0\gamma_{j+1}} - y}{N_0}\right)\right] f_y(y) dy. \tag{17}$$

Now, in order to derive a closed-form expression for (17), we have to solve integrals of the form

$$I = D \int_{t}^{y} \text{erfc}\left(\frac{E-x}{\sqrt{2}C}\right) x e^{-Dx^2} dx. \tag{18}$$

Fortunately, a closed-form solution to (18) can be found in [20, Vol. 2, (1.5.3.10)] as

$$I = g(t, y, z, w, v) = \frac{1}{2}\left\{e^{-t^2 w} - 2e^{-y^2 w} + e^{-t^2 w}\left[1 - \text{erfc}\left(\frac{t-v}{\sqrt{2}z}\right)\right] + \right.$$

$$\left. e^{-y^2 w} \text{erfc}\left(\frac{y-v}{\sqrt{2}z}\right) + \frac{e^{-\frac{wv^2}{1+2z^2 w}}}{\sqrt{1+2z^2 w}}\left[\text{erfc}\left(\frac{t+2tz^2 w - v}{z\sqrt{2+4z^2 w}}\right) - \text{erfc}\left(\frac{y+2yz^2 w - v}{z\sqrt{2+4z^2 w}}\right)\right]\right\}. \tag{19}$$

Finally, combining (19) and (17) a closed-form solution to $\Pi_1$, is derived as

$$\Pi_1 = g\left(\sqrt{N_0}\gamma_j, \sqrt{N_0\gamma_{j+1}}, \sqrt{\frac{N_0}{2}}, \frac{1}{Es\,\Omega}, \sqrt{N_0}\gamma_j\right)$$

$$- g\left(\sqrt{N_0}\gamma_j, \sqrt{N_0\gamma_{j+1}}, \sqrt{\frac{N_0}{2}}, \frac{1}{Es\,\Omega}, \sqrt{N_0\gamma_{j+1}}\right). \tag{20}$$

As a general comment regarding criterion 1, one can observe that $\Pi_1$ may not be always indicative of the efficiency of *S+N* to replace *S/N*, since it involves also the probability that *S/N* leads to a modulation order, $M$. In other words, if the probability that *S/N* leads to the modulation order, $M_j$, is too low, then $\Pi_1$ will be also too low, independently of whether *S+N* led to the same order, $M_j$ or not. Towards this, a supplemental criterion for the efficiency of *S+N* is presented in the following.

*Criterion 2: The probability that the constellation size, determined by S+N, is M, given that the constellation size, determined by S/N is also M,* , i.e.,

$$\Pi_2 = \Pr\left\{ \gamma_j \leq \frac{\left(a\sqrt{E_S} + n_I\right)^2}{N_0} < \gamma_{j+1} \, \middle| \, \gamma_j \leq E_S \frac{a^2}{N_0} < \gamma_{j+1} \right\} \ . \tag{21}$$

Compared to criterion 1, one can observe that this criterion quantifies more efficiently the ability of *S+N* to stand in for *S/N*, since it takes into account the fact that *S/N* has led to a specific modulation order, $M_j$, and not the probability to lead to that specific order.

The probability in (21), can be calculated as in (17), but now, the random variable, $y$, is limited inside the interval $\left[\sqrt{N_0\ \gamma_j}, \ \sqrt{N_0\gamma_{j+1}}\right]$. This is implied by the fact that the probability

$$\Pr\left\{\gamma_j \leq \frac{\left(a\sqrt{E_S} + n_I\right)^2}{N_0} < \gamma_{j+1}\right\} \tag{22}$$

is conditioned on the event that *S/N* leads to the modulation, $M_j$, with probability equal to one, i.e.,

$$\Pr\left\{E_S\frac{a^2}{N_0} \in [\gamma_j,\ \gamma_{j+1})\right\} = 1$$

or equivalently

$$\Pr\left\{y \in \left[\sqrt{N_0\ \gamma_j},\ \sqrt{N_0\gamma_{j+1}}\right)\right\} = 1 \tag{23}$$

Therefore, an approximate solution to (21) can be obtained by "forcing" the pdf, $f_y(y)$, to have unitary integral when integrated from $\sqrt{N_0\ \gamma_j}$ to $\sqrt{N_0\gamma_{j+1}}$. In other words, we have to average the probability

$$\Pr\left\{\sqrt{N_0\gamma_j} - y \leq n_I < \sqrt{N_0\gamma_{j+1}} - y\right\}$$

in (17) over a new pdf, $h_y(y)$, defined as

$$h_y(y) = \frac{f_y(y)}{\int_{\sqrt{N_0\ \gamma_j}}^{\sqrt{N_0\gamma_{j+1}}} f_y(y)dy} \tag{24}$$

so that

$$\int_{\sqrt{N_0\ \gamma_j}}^{\sqrt{N_0\gamma_{j+1}}} h_y(y)dy = 1.$$

Following this approach, (21) can be approximated by

$$\Pi_2 \simeq \Pi_1 \frac{h_y(y)}{f_y(y)}. \tag{25}$$

Numerical results considering these criteria will be presented in the next section, where the *S+N* scheme is applied in a *VR M*-PSK modulation scheme.

## III. VARIABLE RATE *M*-PSK

### A. Mode of Operation

In this section, we introduce a constant power, *VR M*-PSK modulation scheme, in which the decision on the modulation order is not based on the *S/N* samples, but rather on the *S+N* ones. The mode of operation is illustrated in Fig. 1. An ideal coherent phase detection is assumed to be available at the receiver at time instant, $i$, while no channel gain estimation is necessary. The receiver estimates the *S+N* samples at time, $i$, and decides on the constellation size $M_j = 2^j$, $j = 1, 2, \ldots,$ $N$, that will be used at the next time interval, $i + 1$, according to this sample. The decision on the modulation order is sent back to the transmitter, transmitting $N$ bits through a feedback channel, that does not introduce any errors, which can be assured by increasing its delay time and using an ARQ transmission protocol.

As mentioned, the estimation of the modulation order requires a closed-form formula that relates, the *S/N* with the error probability. For the case of the *M*-PSK we can use the approximation for the SER, i.e., [17, Ch. 7]

$$P_M \approx \text{erfc}\left(\sqrt{\gamma}\sin\frac{\pi}{M}\right). \tag{26}$$

Solving (26) with respect to $M$, the maximum constellation size for a given SER is given by

$$M = \frac{\pi}{\arcsin\left(\frac{1}{\gamma}\text{erfc}^{-1}P_M\right)}. \tag{27}$$

Therefore, the selected order, $M$, will be

$$M = M_j, \quad \text{if} \quad \begin{cases} M_j \leq M < M_{j+1} \\ \quad\quad or \\ \gamma_j \leq \frac{\left(a\sqrt{E_S} + n_I\right)^2}{N_0} < \gamma_{j+1} \end{cases}, \tag{28}$$

where

$$\gamma_j = \frac{\text{erfc}^{-1}P_M}{\sin\frac{\pi}{M_j}}. \tag{29}$$

The proposed scheme gives the ability to communications systems with no channel gain estimation capabilities to adapt their transmission rate, so that their spectral efficiency can be increased. We should also note that in some applications, constant envelope modulations, such as *M*-PSK, are more desirable that other ones, because of specific advantages (e.g. their improved performance when used with high power amplifiers).

*B. Performance Evaluation*

*1) Spectral Efficiency:* The normalized spectral efficiency of the *S+N* based *VR M*-PSK scheme is obtained as

$$\mathcal{S}_{S+N} = \frac{R}{B} = \sum_{j=1}^{N} \log_2(M_j) \Pr\left\{ \gamma_j \leq \xi < \gamma_{j+1} \right\}. \tag{30}$$

The probability, $\Pr\left\{ \gamma_j \leq \xi < \gamma_{j+1} \right\}$, in (30) can be rewritten as

$$\Pr\left\{ \gamma_j \leq \xi < \gamma_{j+1} \right\} = \Pr\left\{ \sqrt{\gamma_j N_0} \leq a\sqrt{E_S} + n_I < \sqrt{\gamma_{j+1} N_0} \right\} = \int_{\gamma_j}^{\gamma_{j+1}} f_z(z) dz, \tag{31}$$

where $f_z(z)$ is the pdf of the random variable $z = a\sqrt{E_S} + n_I$, which is the sum of a Rayleigh and a Gaussian random variable, and is given by [15]

$$
\begin{aligned}
f_z(z) &= \frac{1}{\sqrt{2\pi\sigma^2}} \frac{2\sigma^2}{E_s\Omega + 2\sigma^2} e^{-\frac{z^2}{2\sigma^2}} + \frac{z\sqrt{E_s\Omega}}{E_s\Omega + 2\sigma^2} \frac{1}{\sqrt{E_s\Omega + 2\sigma^2}} \\
&\quad \times e^{-\frac{z^2}{E_s\Omega + 2\sigma^2}} \left( 1 - \mathrm{erf}\left( -\frac{\sqrt{E_s}z}{\sqrt{2\sigma^2}\sqrt{E_s\Omega + 2\sigma^2}} \right) \right).
\end{aligned} \tag{32}
$$

A closed-from solution to the indefinite integral

$$\mathcal{J}(z) = \int f_z(z) dz \tag{33}$$

can be found using [20] as

$$
\begin{aligned}
\mathcal{J}(z) &= \frac{2\sigma^2\sqrt{\pi}\,\mathrm{erf}\left(\frac{z}{\sqrt{2\sigma^2}}\right)}{\sqrt{2\sigma^2} + E_s\Omega\sqrt{\pi}} - \frac{e^{-\frac{z^2}{2\sigma^2 + E_s\Omega}}\sqrt{E_s\Omega}}{\sqrt{E_s + 2\sigma^2}\sqrt{E_s\Omega + 2\sigma^2}} \times \\
&\quad \left[ \sqrt{E_s + 2\sigma^2} \left( 1 + \mathrm{erf}\left( \frac{\sqrt{E_s}z}{\sqrt{2\sigma^2}\sqrt{E_s\Omega + 2\sigma^2}} \right) \right) - e^{-\frac{z^2}{2\sigma^2 + E_s\Omega}}\sqrt{E_s}\,\mathrm{erf}\left( \frac{\sqrt{E_s + 2\sigma^2}z}{\sqrt{2\sigma^2}\sqrt{E_s\Omega + 2\sigma^2}} \right) \right].
\end{aligned} \tag{34}
$$

Finally, the spectral efficiency of the *S+N* based *VR M*-PSK scheme is obtained as

$$\mathcal{S}_{S+N} = \sum_{j=1}^{N} \log_2(M_j) \int_{\gamma_j}^{\gamma_{j+1}} f_z(z) dz \tag{35}$$

or

$$\mathcal{S}_{S+N} = \sum_{j=1}^{N} \log_2(M_j) \left[ \mathcal{J}\left(\gamma_{j+1}\right) - \mathcal{J}\left(\gamma_j\right) \right]. \tag{36}$$

On the other hand, the normalized capacity of a *VR M*-PSK scheme, with the modulation order determined by the received *S/N,* is given by

$$\mathcal{S}_{S/N} = \sum_{j=1}^{N} \log_2(M_j) \Pr\left\{ \gamma_j \leq \gamma < \gamma_{j+1} \right\}.$$

For the case that the fading amplitude, $a$, follows the Rayleigh distribution (1), $\gamma$ will be exponentially distributed as

$$f(\gamma) = \frac{1}{\overline{\gamma}} e^{-\frac{\gamma}{\overline{\gamma}}}, \tag{37}$$

where $\overline{\gamma} = \Omega E_s / N_0$ is the average *SNR*. Thus,

$$\Pr\left\{ \gamma_j \leq \gamma < \gamma_{j+1} \right\} = \int_{\gamma_j}^{\gamma_{j+1}} \frac{1}{\overline{\gamma}} e^{-\frac{\gamma}{\overline{\gamma}}} d\gamma = e^{-\frac{\gamma_j}{\overline{\gamma}}} - e^{-\frac{\gamma_{j+1}}{\overline{\gamma}}}$$

and finally, the spectral efficiency of the *S/N* based *VR M*-PSK will be

$$\mathcal{S} = \sum_{j=1}^{N} \log_2(M_j) \left( e^{-\frac{\gamma_j}{\overline{\gamma}}} - e^{-\frac{\gamma_{j+1}}{\overline{\gamma}}} \right)$$

*2) Symbol Error Probability:* Regarding the SER of a *S/N* based *VR M*-PSK scheme, it can be obtained by averaging over the SER for *M*-PSK in AWGN [5], i.e.,

$$P_s = \sum_j \int_{\gamma_j}^{\gamma_{j+1}} P_{AWGN}(M_j, \gamma) f_\gamma(\gamma) d\gamma. \tag{38}$$

The SER of *M*-PSK over AWGN is given by [21, Ch. 8.1]

$$P_{AWGN}(M_j, \gamma) = Q\left(\sqrt{2\gamma}\right) + \frac{2}{\sqrt{\pi}} \int_0^\infty \exp\left\{ -(u - \sqrt{\gamma})^2 \right\} Q\left(\sqrt{2}\gamma \tan\frac{\pi}{M}\right) du \tag{39}$$

The above approach results in some cases to closed-form solutions for the error probability.

For the case of the *S+N*, however, a similar approach cannot be followed, since a formula for the SER as a function of $\xi$ is required. Moreover, when physically realizing *S+N* schemes, though, by sampling the output of a matched filter, the noise is a random variable [22]. Thus, it is inexact to specify the performance of *S+N* systems using a constant noise analysis. Because of the above reasons, the SER for the proposed scheme is calculated via computer simulations.

## C. Application to Equal-Gain Combining Receivers

Consider a multichannel diversity reception system with $L$ branches operating in a discrete-time channel, in which the receiver employs symbol-by-symbol detection. The signal received over the $k$th diversity branch, at the time instant $i$ can be expressed as

$$r_k[i] = \Re\left\{ \left[ a_k[i] e^{j\vartheta_k[i]} s[i] + n_k[i] \right] \right\} = \Re\left\{ R_k[i] \right\}, \quad k = 1\ldots L \tag{40}$$

where $a_k[i]$ is the random magnitude, $\vartheta_k[i]$ is the random phase of the $k$th diversity branch gain and $n_k[i]$, represents the additive noise with $E\left\{ n_k[i]^* n_k[i] \right\} = N_k = 2\sigma^2 = N_0$.

Assuming that the random phase $\vartheta_k[i]$ are known at the receiver, the received signals are co-phased and transferred to baseband so that the signal at the $k$th branch will be

$$u_k[i] = \Re\left\{a_k[i]s(t) + n_k[i]e^{j\vartheta_k[i]}\right\}, \;\; k = 1, \ldots, L. \tag{41}$$

At the combination stage the signals $u_k[i]$ are weighted and summed to produce the decision variable

$$u[i] = \sum_{k=1}^{L} w_k[i]u_k[i] \tag{42}$$

where $w_k[i]$ is the weight of the $k$th branch. Then, applying Maximum Likelihood Detection (MLD), the combiner's output is compared with all the known possible transmitted symbols in order to extract the decision metric.

The coherent equal-gain-combining (EGC) receiver [21, Ch. 9.3] cophases and equally weights each branch before combining and therefore does not require estimation of the channel (path) fading amplitudes, but only knowledge of the channel phase, in order the demodulator to undo the random phase shifts introduced on the diversity channels.

As a result, for $w_k[i] = 1$, the output combined signal will be

$$u[i] = \sum_{k=1}^{L} u_k[i] \tag{43}$$

which is actually the summation of the *S+N* at each diversity branch. As a result, the *VR* M-PSK modulation scheme can be directly applied to EGC receivers with the decision on the modulation order to be based on

$$\psi[i] = \frac{u[i]^2}{N_0} = \frac{\left[\sum_{k=1}^{L} \Re\left\{a_k[i]s(t) + n_k[i]e^{j\vartheta_k[i]}\right\}\right]^2}{N_0} \tag{44}$$

The system model of the EGC receiver employing *VR* M-PSK is shown in Fig. 2.

## IV. RESULTS AND DISCUSSION

The evaluation of the probabilities, $\Pi_1$ and $\Pi_2$ is very important, since they imply the efficiency of *S+N* to estimate the quality of the instantaneous received signal. For the case of the *M*-PSK, $\Pi_1$ and $\Pi_2$ are plotted in Figs. 3 and 4 respectively. Note, that the simulation results coincide perfectly with the analytical ones, for both these probabilities.

A general conclusion derived form these figures is that if the *S/N* based *VR M*-PSK uses a specific modulation order, the probability that the *S+N* based *VR M*-PSK uses the same order is relatively high. Moreover, as expected, the efficiency of the *S/N* improves as the average *S/N* increases, since the instantaneous noise components become small, so that the divergence between $(a\sqrt{E_S}+n)^2$ and $a^2 E_S$ decreases.

The difference between the *S/N* and *S+N* based *VR M*-PSK schemes in terms of spectral efficiency is shown in Fig. 5. It is seen that both schemes achieve the same spectral efficiency, while the difference in their SER is not significant as indicated in Fig. 6. In the latter figure, the SER of the fixed-rate *M*-PSKis also depicted for comparison reasons. In general, it can be observed that the *S+N* can serve as an attractive criterion for determining the channel quality, when channel gain estimation is not available.

The explanation for the efficiency of the *S+N* criterion is as follows: in the conventional *VR M*-PSK, the decision on the modulation order is based on the instantaneous *S/N*, i.e. on $a^2 E_S/N_0$, which means that the instantaneous noise power is ignored, since only the average noise power is taken into account. On the contrary, the *S+N* samples, $(a\sqrt{E_S}+n)^2$ include the instantaneous noise component and therefore an indirect estimation of the instantaneous noise power is obtained.

Similar results can be obtained also for the EGC employing *VR* M-PSK. As shown in Fig.7, the spectral efficiency is increased compared to a fixed rate *M*-PSK scheme, while the SER is maintained under the required levels, whenever this is possible. It should be noted that in some cases the SEP may be greater than the desired target. For example, in Fig. 8, for $L = 4$, we observe that the SEP could be lower than $10^{-3}$ for SNR ¿ 5 dB, by employing BPSK, while with the *VR* MPSK scheme the SEP becomes lower than $10^{-3}$, for SNR ¿ 7 dB.

## V. CONCLUSIONS

A *VR M*-PSK modulation scheme was introduced for wireless communication systems, where channel gain estimation is not available. *VR M*-PSK requires no channel gain estimation and increases the spectral efficiency of *M*-PSK systems under the instantaneous error rate requirement. The choice of the modulation order is not based on the *S/N* samples, but rather on the *S/N* ones. It was analytically shown that the *S+N* criterion is an attractive alternative to *S/N* for choosing the appropriate modulation order in *VR* communication systems, since the same spectral efficiency can be achieved, when either *S/N* or *S+N* is used. Moreover, the proposed adaptive modulation scheme was applied to EGC receivers, enabling low-complexity diversity systems to increase their spectral efficiency.

## VI. FIGURES' CAPTIONS

Fig. 1: The *VR* M-PSK modulation scheme.

Fig. 2: EGC with a *VR* M-PSK modulation scheme.

Fig. 3: The probability, $\Pi_1$, for a *VR* M-PSK modulation scheme. Theoretical (solid lines) and simulation (symbols) results.

Fig. 4: The probability, $\Pi_2$, for a *VR* M-PSK modulation scheme. Theoretical (solid lines) and simulation (symbols) results.

Fig. 5: The spectral efficiency of *VR* M-PSK schemes, for different SER targets.

Fig. 6: The SER of *VR* M-PSK schemes, for different SER targets.

Fig. 7: The spectral efficiency of EGC with *VR* M-PSK schemes, for different SER targets.

Fig. 8: The SER of EGC with *VR* M-PSK schemes, for different SER targets.

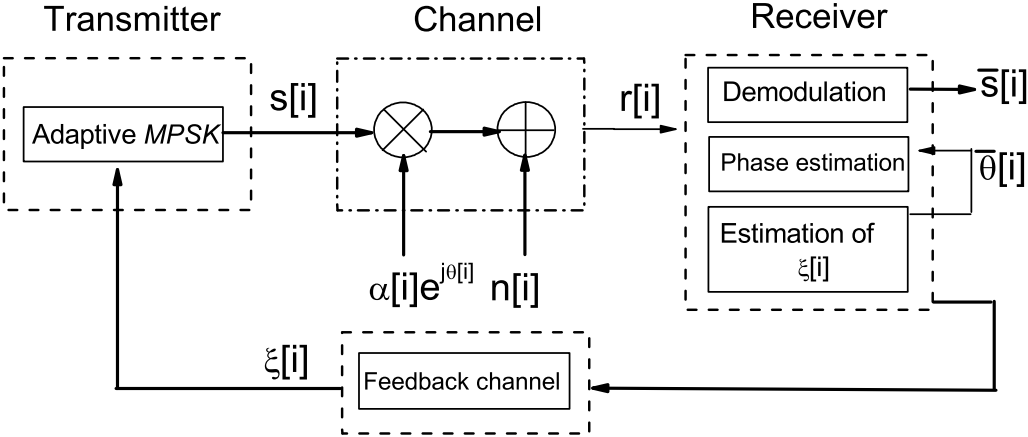

Fig. 1. The *VR* M-PSK modulation scheme.

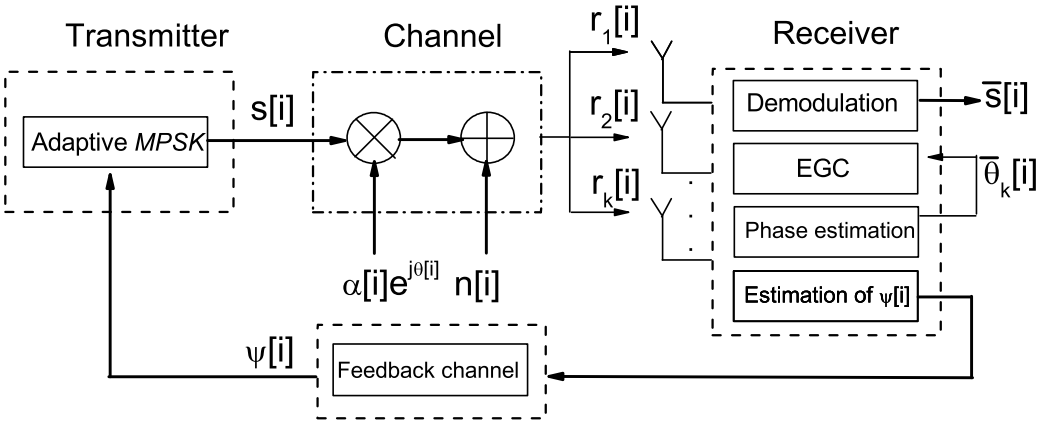

Fig. 2.   EGC with a *VR* M-PSK modulation scheme.

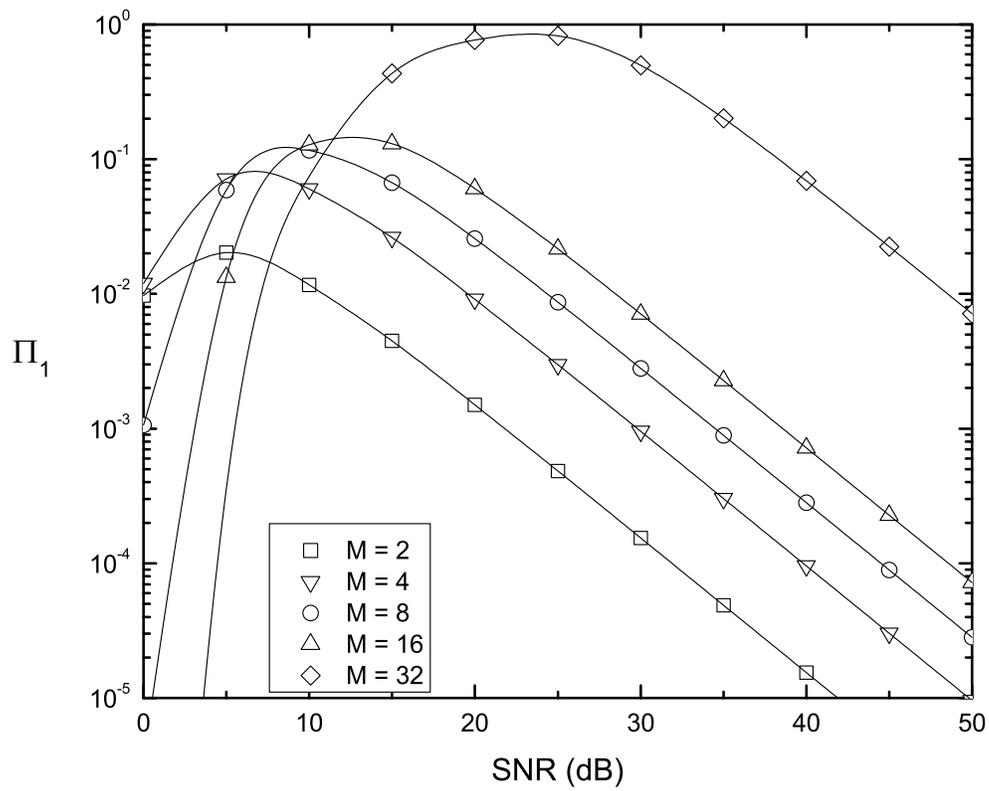

Fig. 3. The probability, $\Pi_1$, for a *VR* M-PSK modulation scheme. Theoretical (solid lines) and simulation (symbols) results.

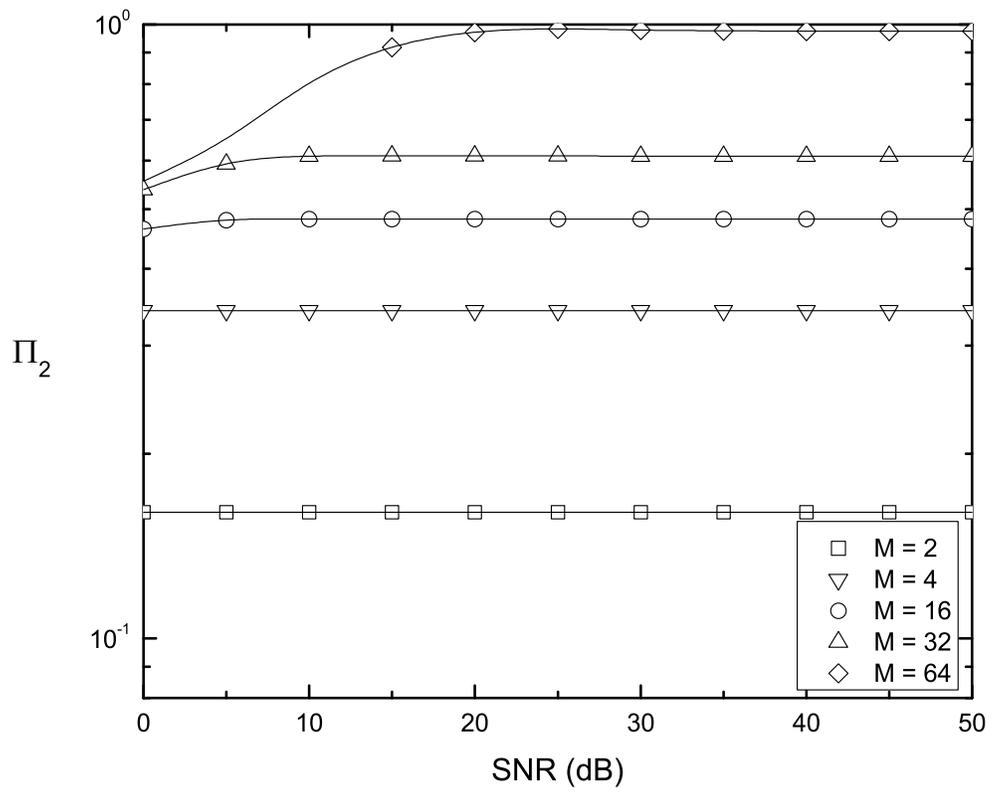

Fig. 4. The probability, $\Pi_2$, for a *VR* M-PSK modulation scheme. Theoretical (solid lines) and simulation (symbols) results.

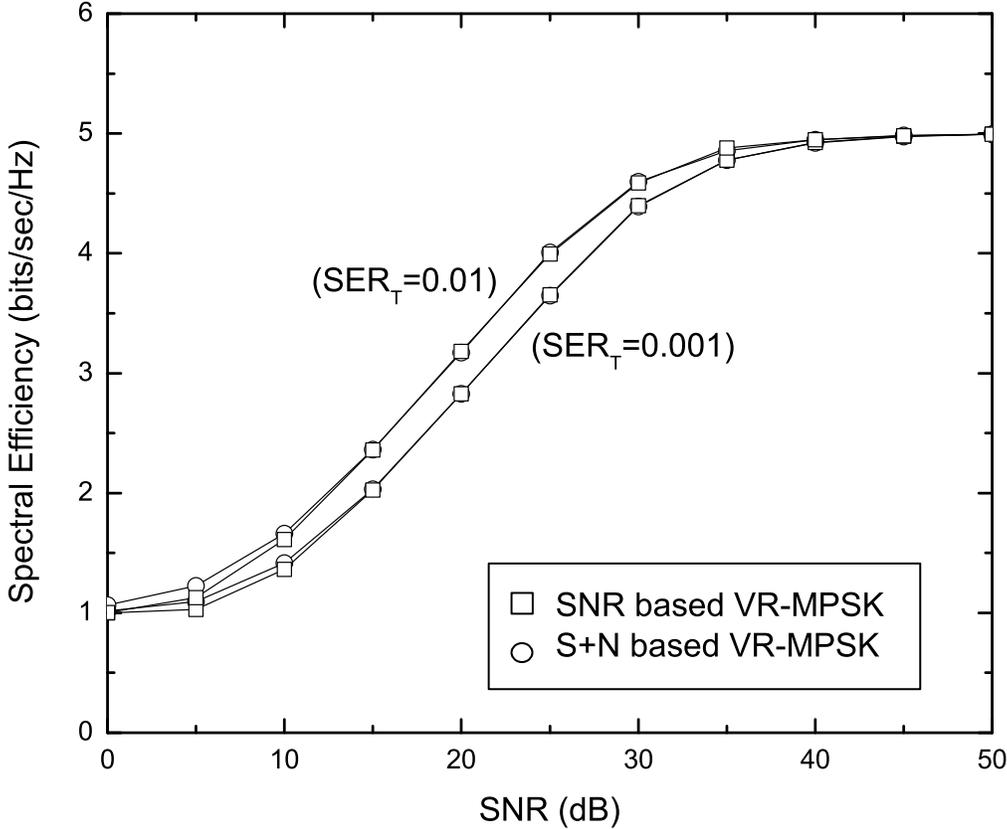

Fig. 5. The spectral efficiency of *VR* M-PSK schemes, for different SER targets.

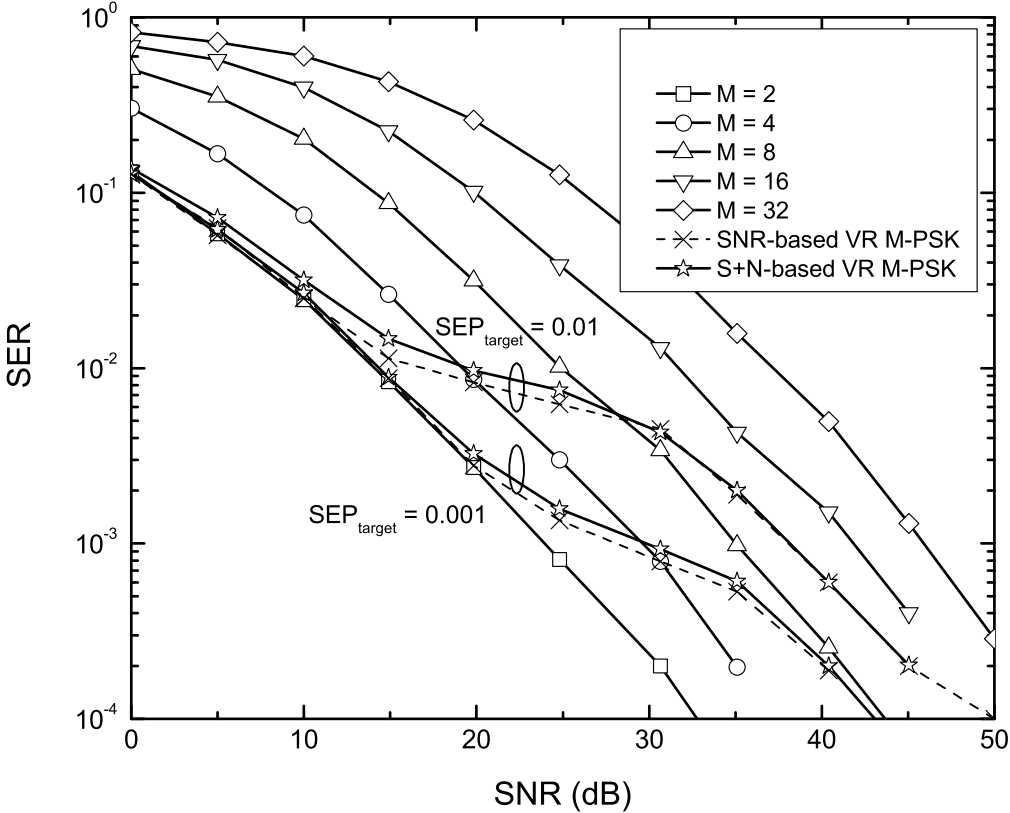

Fig. 6. The SER of *VR* M-PSK schemes, for different SER targets.

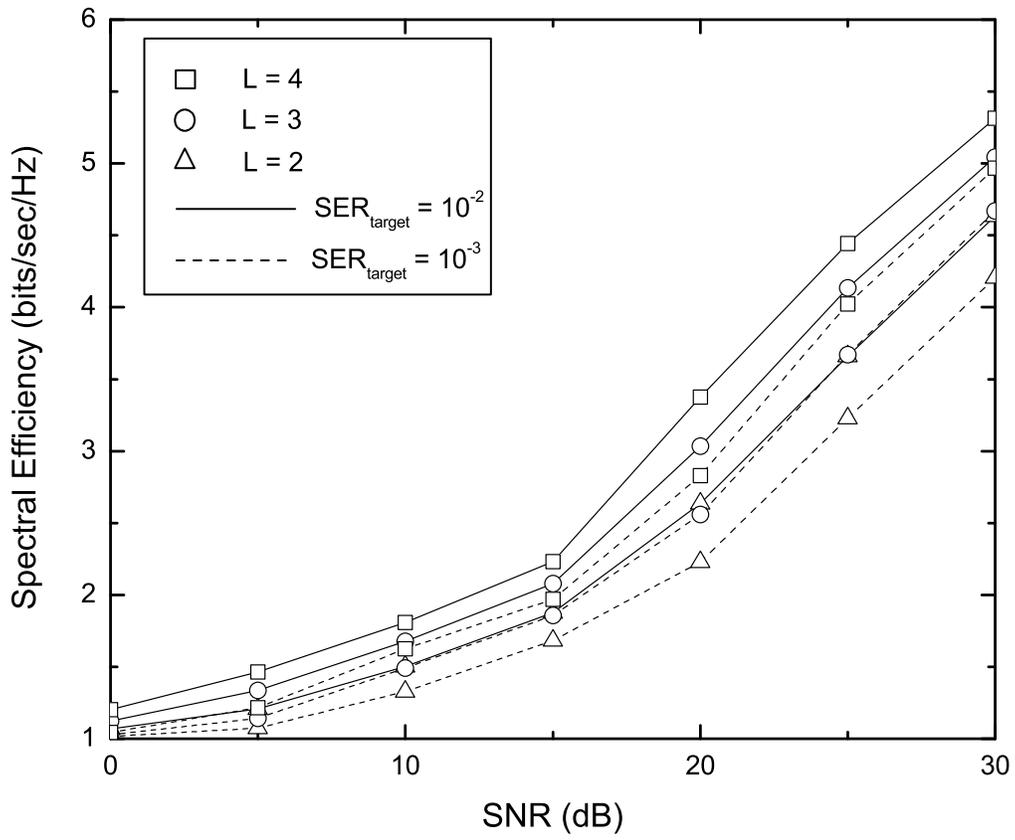

Fig. 7. The spectral efficiency of EGC with *VR* M-PSK schemes, for different SER targets.

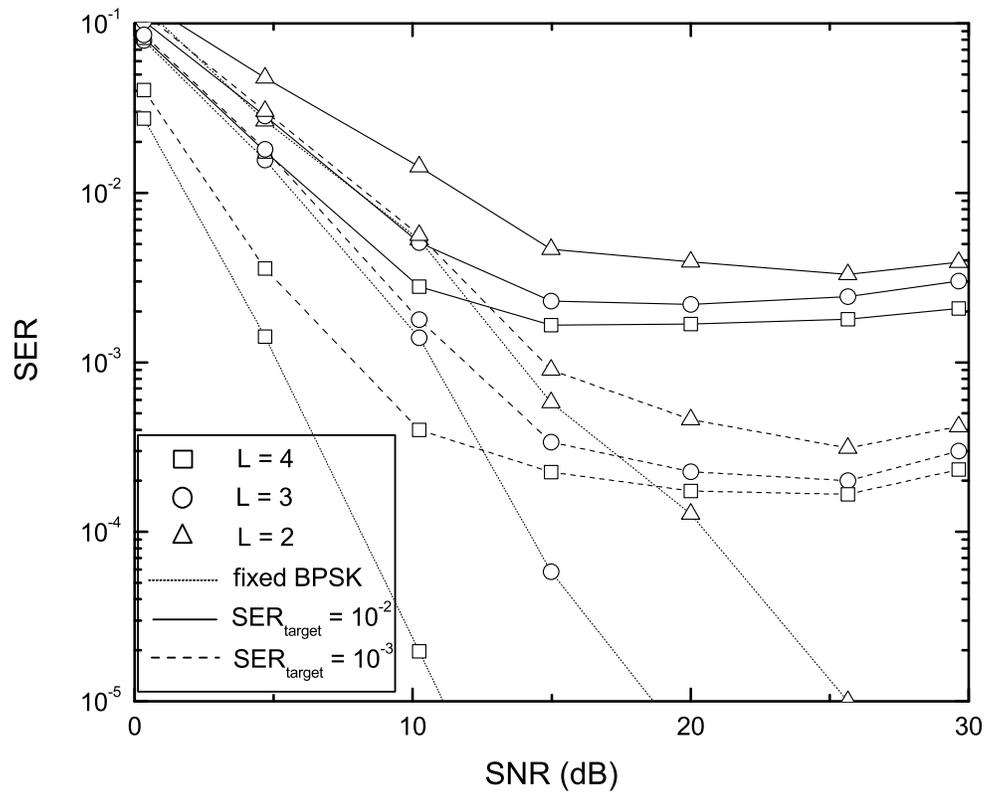

Fig. 8. he SER of EGC with *VR* M-PSK schemes, for different SER targets.